\documentstyle[preprint,aps,epsf,floats,tighten]{revtex} 

\begin{document}

\preprint{\tighten \vbox{\hbox{hep-ph/9810427}
\hbox{} \hbox{} \hbox{} \hbox{} \hbox{} } }

\title{Extracting $|V_{ub}|$ from the inclusive charmless semileptonic 
branching ratio of $b$ hadrons}

\author{Changhao Jin}

\address{School of Physics, University of Melbourne\\
Parkville, Victoria 3052, Australia}

\maketitle

{\tighten
\begin{abstract}%
We calculate the inclusive charmless semileptonic decay width of the $B$ meson
using a QCD-based approach. This approach is able to account for both dynamic
and kinematic effects of nonperturbative QCD. The charmless
semileptonic decay width is found to be enhanced by long-distance strong 
interactions by $+(7\pm 5)\%$ with respect to the free quark decay width. 
Using the resulting theoretical value for the charmless semileptonic 
decay width, an extraction of $|V_{ub}|$ is made from the
measured lifetime and inclusive charmless semileptonic branching ratio of $b$ 
hadrons.
\end{abstract}
}

\newpage

The ALEPH \cite{aleph} and L3 \cite{l3} Collaborations have recently reported 
the first measurements of the
inclusive charmless semileptonic branching ratio of beauty hadrons at LEP. 
Their measurements have been performed exploiting the different
kinematic properties differentiating the rare $b\to X_u\ell\nu$ decay from 
the dominant $b\to X_c\ell\nu$
decay (the symbol $\ell$ indicates either an electron or a muon). The 
inclusive charmless semileptonic branching ratio of beauty
hadrons is measured to be
\begin{eqnarray}
Br(b\to X_u\ell\nu)=&&(1.73\pm 0.55\pm 0.55)\times 
10^{-3}\,\,\,\,\, ({\rm ALEPH}), \nonumber \\
&& (3.3\pm 1.0\pm 1.7)\times 10^{-3}\,\,\,\,\, ({\rm L3}),
\label{eq:branching}
\end{eqnarray}
where in each collaboration the first error is statistical and 
the second systematic.

The measured inclusive charmless semileptonic branching ratio combined with
the measured beauty hadron lifetime can be used to determine the 
Cabibbo-Kobayashi-Maskawa matrix element $|V_{ub}|$, provided the corresponding
inclusive charmless semileptonic decay width except for $|V_{ub}|$ is 
theoretically predicted. This determination of $|V_{ub}|$ would have a smaller
theoretical
error than a determination of it from a partial charmless semileptonic decay
width, since, generally speaking, the latter would be more sensitive to the 
nonperturbative QCD effects. The purpose of this work is to calculate the 
inclusive charmless semileptonic decay
width of the $B$ meson and estimate the theoretical uncertainty in this
calculation, and then using the result to extract $|V_{ub}|$. 

The inclusive charmless semileptonic decay of the $B$ meson is induced by the
underlying $b\to u$ weak transition. However, experimental data are collected 
from decay processes
at the hadron level, which is connected by the nonperturbative QCD interaction
to the underlying quark-level transition. Therefore, in order to calculate
the semileptonic decay width, the nonperturbative QCD effect has to be 
accounted for. 
A theoretical treatment of the nonperturbative QCD effect has been 
developed \cite{jp,jin1,jp1,jin2} on the basis of
the light-cone expansion and the heavy quark effective theory (HQET). 
This approach is from first principles and
the nonperturbative QCD effect can be computed in a systematic way, 
leading to conceptual and practical improvements over the phenomenological 
models of inclusive decays such as the DIS-like parton model \cite{parton} 
and the ACCMM model \cite{accmm}.
Consequently, model independent predictions and a control of theoretical 
uncertainties become possible.

\begin{figure}[t]
\centerline{\epsfysize=9truecm \epsfbox{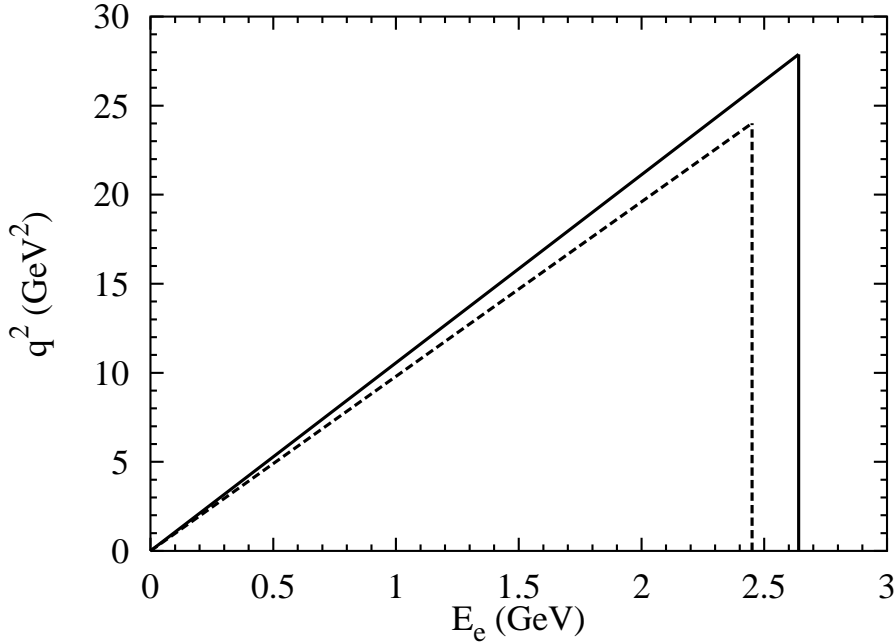}}
\tighten{
\caption[tau1]{Phase space for the $b\to u$ inclusive semileptonic decay.
The interior of the solid curve is the hadron level phase space (the
changeableness of the mass of the final hadronic state is not shown 
explicitly). The interior of the dashed curve is the quark level phase space.} 
\label{fig:phaspa} }
\end{figure}

In this approach the nonperturbative QCD effects in inclusive semileptonic
$B$ decays are ascribed to the distribution function $f(\xi)$, characterizing 
the matrix element of the light-cone bilocal $b$ quark operator between the
$B$ meson states.  The distribution function $f(\xi)$ can be interpreted
as the probability of finding a $b$ quark with momentum $\xi P$ inside the
$B$ meson with momentum $P$.
Nonperturbative QCD contributions to the inclusive semileptonic decay of the
$B$ meson consist of the dynamic and kinematic components.
A crucial observation which emerges from this approach is that  
both dynamic and kinematic effects of  
nonperturbative QCD must be taken into account \cite{jin1}. 
The latter results in the extension of phase space
from the quark level to the hadron level, shown in Fig.~\ref{fig:phaspa}
for the $b\to u$ decay, and turns out to constitute a large part of the 
nonperturbative QCD
contribution to the decay width, which the heavy quark expansion 
approach \cite{hqe} fails to take into account.   

The inclusive charmless semileptonic decay width of the $B$ meson can be 
expressed
as a convolution of the soft nonperturbative distribution function with a
hard perturbative decay rate $\Gamma_{b\to u\ell\nu}$:
\begin{equation}
\Gamma(B\to X_u\ell\nu) = \int_0^1 d\xi\, f(\xi) \Gamma_{b\to u\ell\nu}
(\xi P, \alpha_s)
= \frac{G_F^2M_B^5|V_{ub}|^2}{192\pi^3}
\int_0^1 d\xi\, f(\xi) \xi^5 [1+\sum_{n=1}^\infty 
c_n (\frac{\alpha_s}{\pi})^n],
\label{eq:width}
\end{equation}
where $M_B$ denotes the $B$ meson mass. We have neglected the masses of 
leptons and the $u$ quark.
The interplay between nonperturbative and perturbative QCD effects has been
accounted for since the separation of perturbative and nonperturbative effects
cannot be done in a clear-cut way.
Thus the decay width is expressed in terms of the $B$ meson
mass rather than the $b$ quark mass $m_b$. Equation (\ref{eq:width}) includes
the leading twist contribution; higher-twist contributions are expected to
be suppressed by powers of $M_B$ (or by Sudakov-exponentials).

In the $f(\xi)\to \delta (\xi-m_b/M_B)$ limit, inclusive decay rates are given
by free quark decay.
The first term in Eq.~(\ref{eq:width}) corresponds to the Born-term for free
quark decay. The remaining terms in Eq.~(\ref{eq:width}) contain the 
perturbative QCD corrections, present formally as a series in powers of 
$\alpha_s/\pi$.
The perturbative QCD corrections in Eq.~(\ref{eq:width}) are known to first
order in $\alpha_s$ with the short-distance coefficient \cite{cam}
\begin{equation}
c_1= -\frac{2}{3}(\pi^2-\frac{25}{4}).
\end{equation}
Only the parts of the higher-order perturbative corrections originated from
the running of the strong coupling, the so-called Brodsky-Lepage-Mackenzie
(BLM) corrections \cite{blm},
were calculated to order $\alpha_s^2$ \cite{luke} and to all 
orders \cite{ball}. The BLM corrections are presumably dominant and may
provide an excellent approximation to the full higher-order corrections. 
In our calculation, we include the second-order
BLM correction with the coefficient \cite{luke}
\begin{equation}
c_2^{\rm BLM}= -3.22\beta_0 \, ,
\end{equation}
where $\beta_0=11-2n_f/3$ is the first coefficient of the QCD $\beta$-function.
The second-order BLM correction appears to be rather large. 
However, this does not
mean that the overall $\alpha_s^2$ correction to the charmless
semileptonic decay width is sizeable and a breakdown of perturbation theory.
It was shown \cite{ball,ural} that the large value of the second-order BLM 
correction is solely due to the use of the pole quark mass in the perturbative
calculation. The perturbation series is better behaved when the $B$
decay rate is written by replacing the $b$ quark pole mass with the 
$\Upsilon$ mass \cite{hoang}.

In order to calculate the charmless semileptonic decay width of the $B$ meson,
we need to know the distribution function $f(\xi)$. Due to current 
conservation, it is exactly normalized to unity with a support 
$0\leq\xi\leq 1$. Two additional sum rules for the distribution function can
be obtained by using the operator product expansion and the HQET 
method \cite{hqe}. These two sum rules determine the mean value $\mu$ and the 
variance $\sigma^2$ of the distribution function. The mean value is the 
location of the ``center of mass'' of the distribution function and the 
variance is a measure of the square of its width, which specify the basic form
of the distribution function. The sum rules read \cite{jp,jin1,jp1,jin2}:
\begin{equation}
\mu\equiv\int_0^1 d\xi\, f(\xi)
=\frac{m_b}{M_B}\Bigg (1-\frac{\lambda_1+3\lambda_2}{2m_b^2}\Bigg ),
\label{eq:sum1}
\end{equation}
\begin{equation}
\sigma^2\equiv\int_0^1 d\xi\, (\xi-\mu)^2 f(\xi)
=\frac{m_b^2}{M_B^2}\Bigg [-\frac{\lambda_1}{3m_b^2}-
\Bigg (\frac{\lambda_1+3\lambda_2}{2m_b^2}\Bigg )^2\Bigg ],
\label{eq:sum2}
\end{equation}
where 
\begin{equation}
\lambda_1=\frac{1}{2M_B}<B|\bar h(iD)^2h|B>,
\end{equation}
\begin{equation}
\lambda_2=\frac{1}{12M_B}<B|\bar hg_sG_{\alpha\beta}\sigma^{\alpha\beta}h|B>,
\end{equation}
are the HQET parameters, which parametrize the nonperturbative QCD effects
on a variety of phenomena. The parameter 
$\lambda_2$ can be extracted from hadron spectroscopy:
\begin{equation}
\lambda_2=\frac{1}{4}(M^2_{B^\ast}-M^2_B)=0.12 \ \rm{GeV}^2.
\end{equation}
The parameter $\lambda_1$ suffers from large uncertainty. Fortunately,  from 
the numerical analysis to be discussed below we find that
the result for the charmless semileptonic decay width of 
the $B$ meson is 
insensitive to the variation of $\lambda_1$. For our calculation we adopt the 
value of it from QCD sum rules \cite{bb}:
\begin{equation}
\lambda_1= -(0.5\pm 0.2) \ {\rm GeV}^2.
\label{eq:la1}
\end{equation}
These properties constrain the distribution function considerably.

Nonperturbative QCD methods such as lattice simulation could help determine
further the form of the distribution function. The distribution function could
also be extracted directly from experiment \cite{method}. 
Since these are as yet not done, 
we perform the calculations using the parametrization \cite{jin1} of
the distribution function
\begin{equation}
f(\xi)=N\frac{\xi(1-\xi)^\alpha}{[(\xi-a)^2+b^2]^\beta}\theta(\xi)
\theta(1-\xi),
\label{eq:ansatz}
\end{equation}
where $a, b, \alpha$, and $\beta$ are four parameters, which are constrained 
by the sum rules (\ref{eq:sum1}) and (\ref{eq:sum2}), and $N$ is the 
normalization constant.

The distribution function $f(\xi)$ seems to us not unrelated to the shape 
function introduced in \cite{resum} by resumming
the heavy quark expansion. It would be interesting to discuss the relation
of our approach to theirs.
Their expressions for the decay rate are very similar, although somewhat 
different, to ours obtained in the framework of
light-cone expansion. We can reproduce the formulas of \cite{resum} by 
assuming $\xi=m_b/M_B$. In that case, the scaling variable would be fixed 
to the definite value, instead of distributing in its entire range from 
$0$ to $1$ with the probability $f(\xi)$.  The predictions of these
theoretical approaches are subjected to experimental tests.

\begin{figure}[t]
\centerline{\epsfysize=9truecm \epsfbox{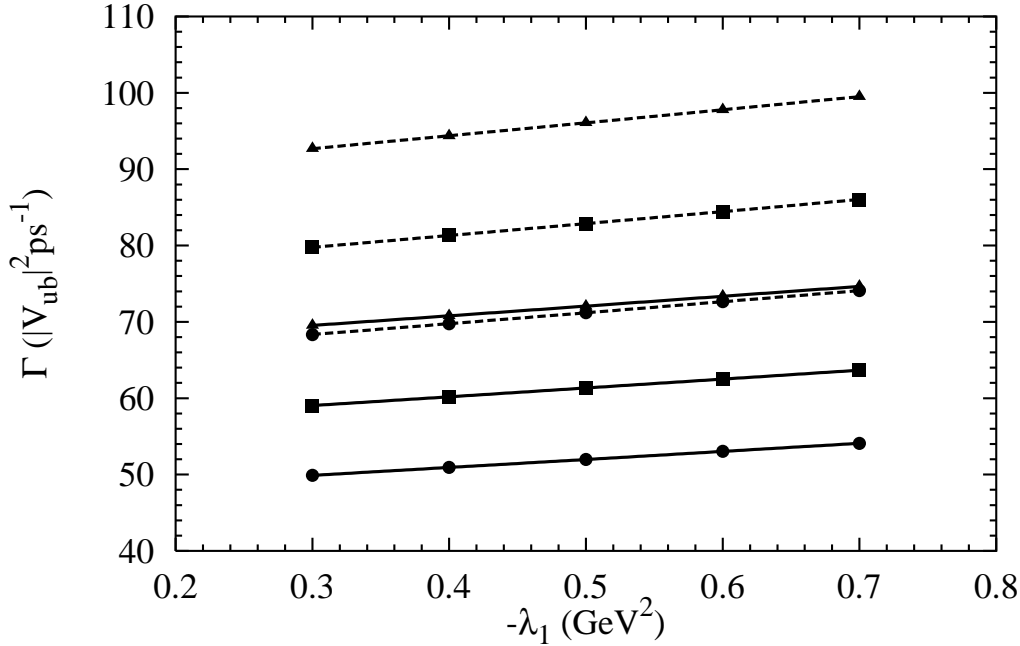}}
\tighten{
\caption[tau1]{Dependence of the charmless semileptonic decay width on the
theoretical input parameters $m_b, \lambda_1,$ and $\mu_r$. The solid (dashed)
curves are for the renormalization scale $\mu_r= m_b/2$ ($\mu_r= 2m_b$). The
curves with solid dots, boxes, triangles correspond to $m_b=4.75, 4.9, 5.05$
GeV, respectively.} 
\label{fig:width_u} }
\end{figure}

Including both the nonperturbative and perturbative QCD 
contributions in the coherent way as described above, 
we are able to calculate the inclusive charmless semileptonic
decay width of the $B$ meson using Eq.~(\ref{eq:width}). 
The decay width is evaluated and the theoretical uncertainties are 
estimated as follows.
\begin{itemize}
\item
We study the variation of the decay width with respect to the mean value and 
the variance
of the distribution function setting $\alpha=\beta=1$ in Eq.~(\ref{eq:ansatz}).
Actually, this amounts
to the study of the decay width as functions of $m_b$ and $\lambda_1$,
since, essentially, the mean value of the distribution function is determined 
by the $b$ quark mass and its variance is determined by $\lambda_1$ according
to the sum rules in Eqs.~(\ref{eq:sum1}) and (\ref{eq:sum2}). The results are
shown in Fig.~\ref{fig:width_u}, varing the $b$ quark pole mass in the range
\begin{equation}
m_b=4.9\pm 0.15 \ {\rm GeV},
\end{equation}
and $\lambda_1$ in the range of (\ref{eq:la1}). The variation of $m_b$ leads
to an uncertainty of $15\%$ in the decay width if other parameters are kept
fixed. A small uncertainty of $4\%$ in the decay width results from the
variation of $\lambda_1$. In other words, the charmless decay width displays
a strong dependence on
the mean value of the distribution function of the $b$ quark 
inside the $B$ meson, but is insensitive to the variance of the 
distribution function.
\item
We examine the further sensitivity of the decay width to the form of the 
distribution 
function when keeping the mean value and variance of it fixed, by varying 
the values of the two additional parameters $\alpha$ and $\beta$ in the
parametrization (\ref{eq:ansatz}). We find that the variation of the decay 
width is typically at the level less than $1\%$ if the form of the 
distribution function
is changed but with the same mean value and variance, hence such a change 
has a negligible impact on the theoretical uncertainty. 
\item
We estimate the uncertainty due to the truncation of the perturbative series
in Eq.~(\ref{eq:width})
by varying the renormalization scale between $m_b/2$ and $2m_b$. We also show
in Fig.~\ref{fig:width_u} the variation of the decay width due to the change 
in the
renormalization scale. We observe that an uncertainty of $15\%$ in the decay 
width stems from the renormalization scale dependence, in agreement with 
the estimate obtained in \cite{ball} based on a different method.
\end{itemize}

This analysis implies that at present the theoretical error has two main
sources: the value of $m_b$ (or equivalently, the mean value of the 
distribution 
function) and the renormalization scale dependence. The  functional 
form of the distribution function, which is not completely known, is likely to
cause some
model dependence. However, the inclusive charmless semileptonic decay width 
of the $B$
meson calculated in this approach is nearly model-independent since
it is only sensitive to the mean value of the distribution function, which is
known from HQET, but insensitive to the detailed form of the distribution
function. It is not surprising that inclusive enough quantities like the total
decay width are less sensitive to nonperturbative QCD effects. Finally, 
adding all the uncertainties in quadrature we find
\begin{equation}
\Gamma(B\to X_u\ell\nu)= (76\pm 16)|V_{ub}|^2 \ {\rm ps}^{-1}.
\label{eq:wvalue}
\end{equation}

\begin{figure}[t]
\centerline{\epsfysize=9truecm \epsfbox{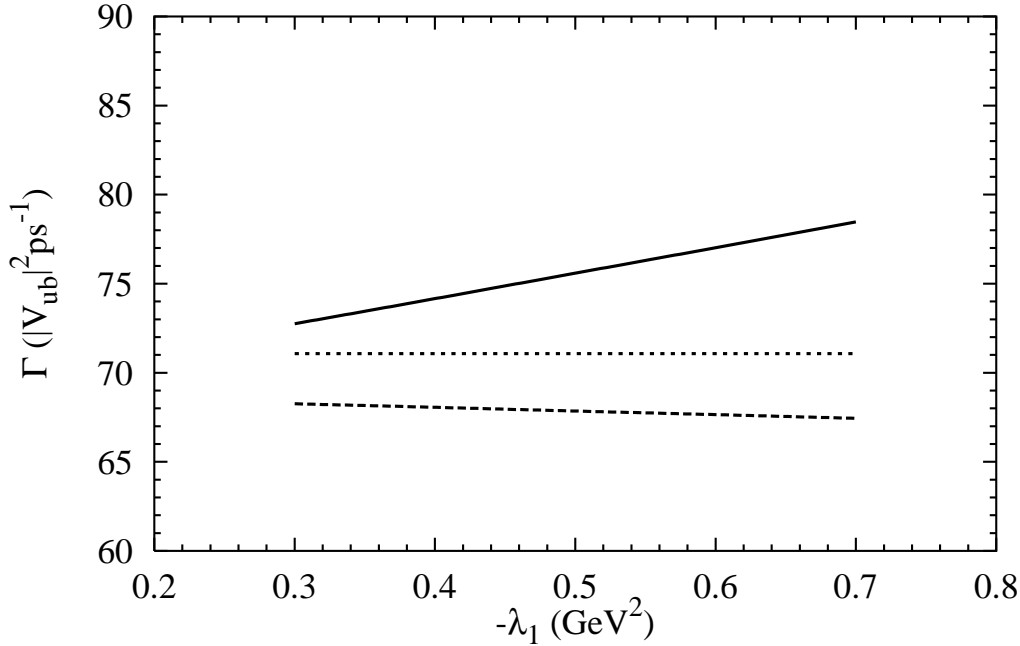}}
\tighten{
\caption[tau1]{Charmless semileptonic decay width as a function of $\lambda_1$
calculated in our approach (solid curve), the free quark decay model (dotted 
curve) and the heavy quark expansion approach (dashed curve). We take $m_b=4.9$
GeV and $\alpha_s=0.2$.} 
\label{fig:fullfree} }
\end{figure}

To see the impact of the nonperturbative QCD effect on the decay width, 
in Fig.~\ref{fig:fullfree} we compare the 
decay widths calculated in our approach, the free quark decay model, and the
heavy quark expansion approach.
The result in our approach shows that the nonperturbative QCD contributions
enhance the decay width by $+(7\pm 5)\%$ with respect to the free quark decay
width, in contrast to the result obtained using the heavy quark expansion 
approach \cite{hqe} (see also \cite{ural}) where
a reduction of the free quark decay width by $-(5\pm 2)\%$ is found.
The enhancement is mainly due to the extension of phase space from
the quark level to the hadron level, indicating the importance of including 
the kinematic nonperturbative QCD effect. This result is in accord with
the finding of a similar calculation of the inclusive $b\to c$ semileptonic 
decay width in \cite{jin1}.

The resulting theoretical value for the charmless semileptonic decay width of 
the $B$ meson in Eq.~(\ref{eq:wvalue}) can be used to determine $|V_{ub}|$. 
Assuming that the average charmless semileptonic decay width of $b$ hadrons is
the same as that of the $B$ meson, using
the average $b$ hadron lifetime $\tau_b=1.564\pm 0.014$ ps \cite{PDG} and
the inclusive charmless semileptonic branching ratio of $b$ hadrons given in
(\ref{eq:branching}) measured by the ALEPH and L3 Collaborations,
we extract
\begin{eqnarray} \label{eq:num}
|V_{ub}|=&& (3.81\pm 0.86\pm 0.40)\times 10^{-3}\,\,\,\,
({\rm ALEPH}),\nonumber \\
&& (5.27\pm 1.60\pm 0.55)\times 10^{-3}\,\,\,\, ({\rm L3}),
\end{eqnarray}
where in each case the first error is experimental and the second theoretical. 

In conclusion, we have extracted $|V_{ub}|$ from the inclusive charmless 
semileptonic branching ratio of $b$ hadrons measured recently by the ALEPH and
L3 Collaborations. The inclusive charmless semileptonic decay 
width of the $B$ meson has been calculated in a QCD-based approach with 
theoretical uncertainties under control. This approach is able to include both
dynamic and kinematic components of the nonperturbative QCD effects. We have
shown that including the latter is crucial, such that the charmless decay 
width is enhanced, which leads to a decrease in the extracted value for 
$|V_{ub}|$. It is encouraging that the nearly model-independent
determination of $|V_{ub}|$ from the
inclusive charmless semileptonic branching ratio of $b$ hadrons is consistent
with those determined by measuring the lepton energy spectrum above the 
endpoint of the inclusive $b\to c\ell\nu$ spectrum or by measuring the 
branching ratios for the exclusive decays $B\to\pi\ell\nu$ and 
$B\to\rho\ell\nu$ \cite{PDG}.  
  
Currently, the error bar associated with $|V_{ub}|$ determined from the 
inclusive charmless semileptonic branching ratio of $b$ hadrons is 
not predominantly theoretical in origin. Both experimental and theoretical
efforts are needed to reduce the error. We observe that
the uncertainty from perturbative QCD is already one of the major theoretical 
limiting factors, comparable in size to another one, i.e., the uncertainty 
from nonperturbative QCD. 
To reduce the error due to the perturbative QCD correction, a complete 
calculation to order $\alpha_s^2$ is necessary.
Moreover, to reduce the theoretical uncertainty from nonperturbative QCD
requires more accurate knowledge of the mean value of the distribution function
(or the $b$ quark mass). Higher-twist contributions, 
which are not included in our calculation, also deserve investigation,
which may still yield numerically sizable effects of order 
$\Lambda_{\rm QCD}/M_B \sim 5\%$.

In order to extract $|V_{ub}|$, the overwhelming ``background'' of $b\to c$
transitions necessitates the introduction of experimental cuts if statistical
errors are to be minimized.
The hadronic invariant mass spectrum appears to be a promising observable for
discriminating $b\to u$ signal from $b\to c$ background \cite{mx}.
We believe that in the future a
measurement of the differential decay rate $d\Gamma(B\to X_u\ell\nu)/d\xi_u$
will be very useful for a high precision $|V_{ub}|$ determination with
a minimum overall (theoretical and experimental) error.   
It has been shown \cite{method} that this differential decay rate is 
proportional to the nonperturbative distribution function.
The measurement of the decay distribution with respect to the scaling
variable $\xi_u$ is of particular significance in a range of issues.
They include \cite{method}:
\begin{itemize}
\item 
The $B\to X_u\ell\nu$ decay can be differentiated from the
$B\to X_c\ell\nu$ decay even more efficiently than the cut on the hadronic
invariant mass. 
\item
A model-independent determination of $|V_{ub}|$ can be obtained by avoiding 
hadronic complications.
\item
Improved measurement of the inclusive charmless semileptonic branching ratio 
of $b$ hadrons can be achieved.
\item
The nonperturbative distribution function can be extracted directly.
\end{itemize}
It remains to be seen whether such a measurement is feasible.

\acknowledgments
This work was supported by the Australian Research Council.

{\tighten

} 

\end{document}